\documentclass[11pt]{article}
\usepackage{graphicx}

\begin{document}

\def\xslash#1{{\rlap{$#1$}/}}
\def \p {\partial}
\def \dd {\psi_{u\bar dg}}
\def \ddp {\psi_{u\bar dgg}}
\def \pq {\psi_{u\bar d\bar uu}}
\def \jpsi {J/\psi}
\def \psip {\psi^\prime}
\def \to {\rightarrow}
\def\bfsig{\mbox{\boldmath$\sigma$}}
\def\DT{\mbox{\boldmath$\Delta_T $}}
\def\xit{\mbox{\boldmath$\xi_\perp $}}
\def \jpsi {J/\psi}
\def\bfej{\mbox{\boldmath$\varepsilon$}}
\def \t {\tilde}
\def\epn {\varepsilon}
\def \up {\uparrow}
\def \dn {\downarrow}
\def \da {\dagger}
\def \pn3 {\phi_{u\bar d g}}

\def \p4n {\phi_{u\bar d gg}}

\def \bx {\bar x}
\def \by {\bar y}

\begin{center}
{\Large\bf Inconsistences of the $k_T$-Factorization in Exclusive B-Meson Decays}
\vskip 10mm
F. Feng$^1$, J.P. Ma$^1$ and Q. Wang$^{2}$    \\
{\small {\it $^1$ Institute of Theoretical Physics, Academia Sinica,
Beijing 100190, China }} \\
{\small {\it $^2$ Department of Physics and Institute of Theoretical
Physics, Nanjing Normal University, Nanjing, Jiangsu 210097, P.R.China}} \\
\end{center}
\vskip 10mm

\begin{abstract}
The $k_T$-factorization has been widely used for exclusive decays of
$B$-mesons. In this factorization the pertubative coefficients
are extracted from scattering of off-shell partons. Because
the off-shellness of partons the extracted pertubative coefficients
in general are not gauge-invariant. We show that these
perturbative coefficients contain gauge-dependent singularities
beyond tree-level. This leads to that the $k_T$-factorization
is gauge-dependent and violated
in the general covariant gauge. This is not consistent with expectations of physics.
In the factorization
there exist
two different wave functions for a $B$-meson in general.
We show that the two wave functions satisfy different evolution equations
whose solutions are used to re-sum large log's. Based on this fact,
the rseummation of large log's at leading log approximation in
the widely used $k_T$-factorization with only one wave function
is inconsistent.
\end{abstract}

\par\vskip30pt

\par
Studies of exclusive $B$-meson decays in experiment and theory play an important
role in testing the standard model. For those decays involving large momentum
transfers two theoretical approaches exist. One of them is
based on the
collinear factorization\cite{BBNS}, in which the transverse
momenta of partons in a hadron are integrated out and their
effect at leading twist is neglected. The collinear factorization
has been  proposed for other exclusive processes for long
time\cite{BL,CZrep}. Another one is based on the $k_T$-factorization\cite{KT1},
where one takes the
transverse momenta of partons into account at leading twist by
meaning of  wave functions. The advantage of the $k_T$-factorization is that
it may eliminate end-point singularities in
collinear factorization\cite{KT1} through resummation of large log's and some higher-twist effects
are included. The $k_T$-factorization, also called as pQCD approach, has been widely used\cite{KTB1,KTB2}
and it is quite successful in phenomenology. However, the $k_T$-factorization
has not been examined so far beyond tree-level.
\par
In the $k_T$-factorization for exclusive $B$-decays the perturbative
coefficients are extracted
from scattering amplitudes of off-shell partons. Therefore one can not expect
in general that these perturbative coefficients, hence the factorization, are gauge-invariant.
Since the $k_T$-factorization
has not been examined beyond tree-level for any exclusive $B$-decay,
the question if the factorization is really gauge invariant is not answered.
Using the $k_T$-factorization for processes involving light hadrons only,
e.g., the transition $\pi \gamma^* \to \gamma$,
one can show that the one-loop perturbative coefficient contains gauge-dependent
singularities\cite{FMW}. The singularities are light-cone singularities and come only from
the wave function of $\pi$. The singularities do not appear in Feynman gauge\cite{FMW,NLi}.
This leads to the conclusion that the $k_T$-factorization is in general gauge-dependent
and violated. The purpose of the current work is to examine this issue of exclusive decays
of a $B$-meson. It should be noted that in the so-called transverse momentum dependent factorization\cite{MWF},
perturbative coefficients are extracted from scattering amplitudes of on-shell partons,
hence they are gauge-independent.
\par
Another problem of the $k_T$-factorization is related to the wave functions of $B$-mesons.
In the heavy quark limit, there exist in general two wave functions, while
in the $k_T$-factorization only one wave function is employed\cite{KT1,KTB1,KTB2}.
This leads to the question if the two wave functions are the same\cite{DeSa}.
There is no exact proof for the equivalence of the two wave functions.
In fact, as we will show, the two wave functions satisfy different evolution
equations. This clearly indicates that the two wave functions are not same.
The mentioned evolution equations have the usage for the resummation
of large logarithms terms in perturbative coefficients. Because the two wave functions
and their evolution equations are different, the resummation
of large logarithms terms in the $k_T$-factorization with only one wave function
can not be consistent.
\par
We take the semi-leptonic decay $B\to \pi \ell \nu_\ell$ as an example
to study the above problems. We will use Heavy Quark Effective Theory(HQET)
for the heavy $b$-quark. Its field in HQET is $h$. The field $h$ depends on the
velocity $v$ with which the $B$-meson moves. In the heavy quark limit,
the $B$-meson mass $m_B$ is the same as the $b$-quark mass $m_b$, hence
we have the $B$-meson momentum $P =m_B v \approx m_b v$ with $v^2=1$.
The decay amplitude is determined by the matrix element,
which can be decomposed into two form factors:
\begin{eqnarray}
 \langle \pi (K) \vert \bar q \gamma ^\mu h\vert \bar B(v)\rangle &=&
 F_+ (q^2) (P+K)^\mu + \left ( m_B^2 - m^2_\pi \right )\frac{ q^\mu}{q^2}
   \left [ F_0(q^2) - F_+ (q^2) \right ]
 \nonumber\\
      &=& F_P (q^2) P^\mu +  F_K (q^2) K^\mu.
\end{eqnarray}
In the above $q$ is the momentum transfer $q =P-K$. The two form factors
$F_{0,+}$ are standard. We introduce two form factors $F_{P,K}$ which are linear combinations
of the standard two. The implication of $F_{P,K}$  will become clear later.
\par
\par

\begin{figure}[hbt]
\begin{center}
\includegraphics[width=15cm]{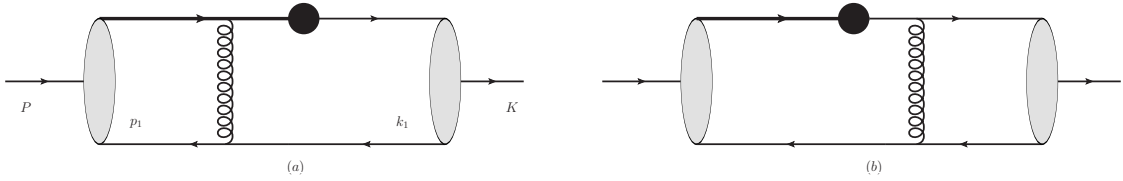}
\end{center}
\caption{The leading order contribution. The thick line is for $b$-quark in HQET.  }
\label{Feynman-dg1}
\end{figure}
\par
The leading order result of the $k_T$-factorization can be derived from the contributions
in Fig.1.
We take a frame in which the $\bar B$-meson moves in the $z$-direction. We will use the  light-cone coordinate system, in which a
vector $a^\mu$ is expressed as $a^\mu = (a^+, a^-, \vec a_\perp) =
((a^0+a^3)/\sqrt{2}, (a^0-a^3)/\sqrt{2}, a^1, a^2)$ and $a_\perp^2
=(a^1)^2+(a^2)^2$.
We also introduce two light-cone vectors: $n^\mu =(0,1,0,0)$ and $l^\mu = (1,0,0,0)$.
In the light-cone coordinate system $v$ is given by $v^\mu=(v^+,v^-,0,0)$.
In the first step, the matrix element based on Fig.1 can be written as:
\begin{eqnarray}
 \langle \pi (K) \vert \bar q \gamma^\mu h \vert \bar B(v)\rangle &=& \frac{N_c^2-1}{4 N_c^2} \int dk_1^- d^2k_{1\perp}d^4 p_1
  H^\mu_{\alpha\beta \gamma\rho} (p_1,k_1)
                 \left (\gamma_5  \gamma^+ \right )_{\rho \gamma } \phi_{\pi}(y,k_{1\perp})
\nonumber\\
   && \cdot \int \frac {d^4 x}{(2\pi)^4} e^{ip_1\cdot x} \langle 0 \vert \bar q_\beta (x) (h_v)_\alpha (0) \vert \bar B(v) \rangle,
\nonumber\\
   H^\mu_{\alpha\beta \gamma\rho}(p_1,k_1) &=& \frac{-i}{(p_1-k_1)^2} \left [ -i g_s \gamma_\nu \right ]_{\beta \rho}
        \left \{ \left [ \gamma^\mu \frac{i}{-v\cdot k_1 +i\varepsilon} (-ig_s v^\nu )\right ]_{\gamma\alpha}
\right.
\nonumber\\
   && \left. \ \ \ \
        + \left [ (-i g_s \gamma^\nu) \frac{i\gamma\cdot (K-p_1)}{(K-p_1)^2+i\varepsilon} \gamma^\mu \right ]_{\gamma\alpha}\right \},
\end{eqnarray}
where we have already evaluated out the trivial color factors. $\alpha,\beta,\gamma$ and $\rho$ are Dirac indices. We have also
made the leading twist expansion for the quark density matrix element of $\pi$ and taken the leading order result with the
wave function $\phi_\pi(y,k_{1\perp})$. Hence the momentum $k_1$ is given by $k_1^\mu =(0, y K^-, \vec k_{1\perp})$.
In the $k_T$ factorization one sets $p_1^-=0$ in $H^\mu_{\alpha\beta \gamma\rho} (p_1,k_1)$, the integral over $p_1^-$
and $z^+$ can be then performed directly. One has
\begin{eqnarray}
 \langle \pi (K) \vert \bar q \gamma^\mu h \vert \bar B(v)\rangle &=& \frac{N_c^2-1}{4 N_c^2} \int dk_1^- d^2k_{1\perp}d p_1^+ d^2 p_{1\perp}
  H^\mu_{\alpha\beta \gamma\rho} (p_1,k_1)\vert_{p_1^-=0}
                 \left ( \gamma_5 \gamma^+ \right )_{\rho \gamma } \phi_{\pi}(y,k_{1\perp})
\nonumber\\
   && \cdot \int \frac {d x^- d^2 x_\perp}{(2\pi)^3} e^{ip_1^+ x^- -i\vec p_{1\perp}\cdot \vec x_\perp}
   \langle 0 \vert \bar q_\beta (x) h_\alpha (0) \vert \bar B(v) \rangle \vert_{x^ + =0}.
\end{eqnarray}
The above quark density matrix of the $B$-meson can be decomposed with the $16$ $\Gamma$-matrices combined
with the vector $v^\mu$, $p_{1\perp}^\mu$ and $n^\mu$ or $l^\mu$. The appearance
of the vector $n$ or $l$ here is because a light-cone direction in the density matrix is chosen.
It should be noted that the two quark fields are not separated along a light-cone in the space-time.
The decomposition can be written as:
\begin{eqnarray}
 \Gamma_{\alpha\beta } ( k^+, k_\perp) &=& \int\frac{d z^- d^2 z_\perp}{(2\pi)^3}
 e^{i k^+ z^- -i\vec k_\perp \cdot \vec z_\perp }
 \langle 0 \vert \bar q_\beta (z) h_\alpha (0) \vert \bar B(v) \rangle \vert_{z^+ =0}
\nonumber\\
  &=& -\left [ \frac{ \gamma \cdot v + 1}{4}
        \left (   \gamma^- \phi_+(k^+, k_\perp) + \gamma^+ \frac{v^- }{v^+} \phi_- (k^+, k_\perp)
         + \cdots \right )\gamma_5 \right ]_{\alpha\beta},
\end{eqnarray}
where the $\cdots$ stand for the terms involving $\gamma_\perp^\mu$.
The contraction of $\gamma_\perp$ with $H$ will produce perturbative coefficients related
to these terms and the perturbative coefficients are suppressed by an extra power of $p_{1\perp}/m_b$
or $k_{1\perp}/m_b$ in comparison with that related to $\phi_+$ or $\phi_-$.
In the leading power we can neglect these terms. From the above one has two wave functions for $B$-meson.
The exact definitions of the wave functions $\phi_+$, $\phi_-$ and $\phi_\pi$ will be given later.
With these wave functions
we obtain then the leading order result in the $k_T$-factorization for the form factors:
\begin{eqnarray}
F_P (q^2) P^+  &=& m_b v^+ \int  dx dy d^2k_{\perp}d^2 p_{\perp}  \phi_+(x, p_{\perp}) \phi_\pi (y, k_{\perp})
         H_P (x,\vec p_\perp, y, \vec k_\perp),
\nonumber\\
F_K (q^2) K^-  &=&
 m_b v^- \int dx dy d^2k_{\perp} d^2 p_{\perp}
 \phi_-(x, p_{\perp}) \phi_\pi (y, k_{\perp}) H_K(x,\vec p_\perp, y, \vec k_\perp),
\end{eqnarray}
with the perturbative coefficients
\begin{eqnarray}
   H_P (x,\vec p_\perp, y, \vec k_\perp) &=&
4\pi \alpha_s   \frac{N_c^2-1}{2 N_c^2} \left [
 \frac{x \hat m_b^2 }{ \left (x y \hat m_b^2 +(\vec p_{\perp} -\vec k_{\perp})^2 \right )
 \left ( x \hat m_b^2 + p^2_{\perp}\right ) } \right ] + {\mathcal O}(\alpha_s^2),
\nonumber\\
H_K(x,\vec p_\perp, y, \vec k_\perp) &=& 4\pi \alpha_s   \frac{N_c^2-1}{2 N_c^2} \left [
 \frac{1}{ y  \left ( x y \hat m_b^2 +(\vec p_{\perp} -\vec k_{\perp})^2 \right ) } \right ]
  + {\mathcal O}(\alpha_s^2) ,
\nonumber\\
   x &=& \frac{p_1^+}{P^+} = \frac{p_1^+}{m_b v^+}, \ \ \ \ y=\frac{k_1^ -}{K^-}, \ \ \ \ \hat m_b^2 =m_b^2 - q^2.
\end{eqnarray}
The integration range is given by $0\leq  x < \infty$ and $ 0 \leq y \leq 1$.
\par
From the derivation of the leading order results several observations can be made:
If we set $\phi_+ = \phi_-$, then the above results reduces to that of \cite{KT1}
for the $k_T$-factorization of the decay in the heavy quark limit,
where only one wave function for the $B$-meson is employed. We will show later that the two wave functions
can not be same. In the above $k_T$-factorization, the partons entering the hard scattering are off-shell.
The initial light antiquark
has the momentum $p^\mu =(xP^+,0,\vec p_\perp)$ and the $b$-quark in HQET carries the momentum $-p$.
The outgoing antiquark carries the momentum $k^\mu =(0,yK^-, \vec k_\perp)$.
This implies that one should also take these off-shell partons to calculate
the higher order corrections of $H_{P,K}$, as illustrated in \cite{NLi}.
A part of results of the collinear factorization can be obtained from the above results by
replacing the wave functions with the corresponding light-cone wave functions  and
discard the $k_T$-dependence in the perturbative coefficients and the $k_T$-integrals.
Then we see that the so called end-point singularity appears in $F_K$ with $y\to 0$,
but not in $F_P$.
\par
The above is derived at the tree-level in Feynman gauge.
In fact, the perturbative coefficients are already gauge-dependent
at tree-level\cite{Wei} because the scattering is of off-shell partons.
We will examine the factorization beyond the tree level. Following the above
derivation we need to calculate the form factors and the wave functions
by using the parton states with the same off-shell momenta and
the corresponding projections for external legs indicated in Eq.(2,4) beyond the tree level.
Then we can extract the perturbative coefficients via:
\begin{eqnarray}
F_P^{(1)} &=&
   H_P^{(1)}\otimes \phi_+^{(0)} \otimes\phi_\pi^{(0)} + H_P^{(0)}\otimes \phi_+^{(1)} \otimes\phi_\pi^{(0)}
   +H_P^{(0)} \otimes\phi_+^{(0)}\otimes \phi_\pi^{(1)},
\nonumber\\
\frac{K^-}{P^-} F_K^{(1)} &=&
   H_K^{(1)}\otimes \phi_-^{(0)} \otimes\phi_\pi^{(0)} + H_K^{(0)}\otimes \phi_-^{(1)} \otimes\phi_\pi^{(0)}
   +H_K^{(0)} \otimes\phi_-^{(0)}\otimes \phi_\pi^{(1)}.
\end{eqnarray}
In this paper we will denote the $n$-loop contribution to a quantity $A$ as $A^{(n)}$.
From the above, one can see that the one-loop contribution $H_{P,K}^{(1)}$ receives
contribution from the form factor and also from the wave functions.
\par
We first study the one-loop contributions from wave functions.
The two wave functions of $B$-meson can be consistently defined by supplying gauge links. We introduce:
\begin{equation}
L_u (\infty, z) = P \exp \left ( -i g_s \int_0^{\infty} d\lambda
     u\cdot G (\lambda u + z) \right ) .
\end{equation}
The wave functions can be defined by taking the limit $u^-\gg u^+$\cite{LiLiao,MWWF}:
\begin{eqnarray}
 \phi_{+,-}(k^+, k_\perp,\zeta) &=& \int \frac{ d z^- d^2 z_\perp }{(2\pi)^3}
  e^{ik^+z^- - i \vec z_\perp\cdot \vec k_\perp}
 \langle 0 \vert \bar q(z) L_u^\dagger (\infty, z)
  \left (\gamma^+, \frac{v^+}{v^-} \gamma^- \right ) \gamma_5 L_u (\infty,0) h(0) \vert \bar B(v) \rangle,
\nonumber\\
  \zeta^2 &=& \frac{2 u^- (k^+)^2}{u^+}\approx
\frac{ 4 (u\cdot k )^2}{u^2},
\end{eqnarray}
with $z^\mu =(0,z^-,\vec z_\perp)$ and $k^+ = x P^+$. The wave functions also depend on $\mu$, the renormalization
scale, and on $\zeta^2$. The dependence of $\zeta^2$ will be used for resummation.
The limit $u^- \gg u^+$ should be understood as one neglects any contribution which is proportional
to any positive power of $u^+$.
To calculate the wave functions consistently with the
factorization, we replace the $\bar B$ meson with a quark pair $\bar q(p) b(-p)$. The quarks are
off-shell, and $p$ is given by
$p^\mu = (p^+, 0, \vec p_\perp)$. At tree-level one simply has:
\begin{equation}
 \phi_{+}^{(0)}(k^+, k_\perp) = \delta(k^+ -p^+) \delta^2 (\vec k_\perp -\vec p_\perp), \ \ \ \
 \phi_{-}^{(0)}(k^+, k_\perp) = \delta(k^+ -p^+) \delta^2 (\vec k_\perp -\vec p_\perp).
\end{equation}
\par
To examine the gauge dependence
we will use the general covariant gauge to calculate the one-loop corrections to the wave functions.
In this gauge the gluon propagator takes the form
\begin{equation}
  \frac{-i }{q^2 + i\varepsilon} \left ( g^{\mu\nu} - \alpha \frac{q^\mu q^\nu}{q^2 + i\varepsilon} \right ).
\end{equation}
The Feynman gauge is obtained by taking $\alpha=0$. We will call the $\alpha$-dependent corrections
as gauge parts and denote these corrections with the sub-index $\alpha$.
At one-loop the corrections come from diagrams given in Fig.2 and Fig.3.
\par

\begin{figure}[hbt]
\begin{center}
\includegraphics[width=9cm]{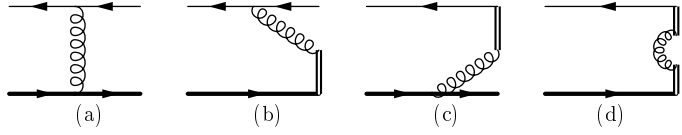}
\end{center}
\caption{The one-loop contributions to wave functions. }
\label{Feynman-dg1}
\end{figure}
\par
As mentioned at the beginning, the one-loop corrections will contain the so called light-cone singularities.
We give a detailed calculation of Fig.2c as an example to illustrate the appearance of light-cone singularities.
The contribution from Fig.2c to $\phi_{+}$ and $\phi_{-}$ is:
\begin{eqnarray}
\phi_\pm (k^+, k_\perp) \vert_{2c}  &=& \int \frac{d^4 q }{(2\pi)^4} \delta(k^+ -p^+ + q^+)
\delta^2 (\vec k_\perp - \vec p_\perp + \vec q_\perp ) \frac{i} {v\cdot (q-p)+ i\varepsilon}
\frac{-i}{-u\cdot q -i\varepsilon}
\nonumber\\
 && \cdot {\rm Tr} \left [ -\frac{1+\gamma\cdot v}{4 N_c} \gamma^{\mp} \gamma_5 \gamma^{\pm} \gamma_5
   (ig_s u^\mu T^a) (-i g_s v^\nu T^a) \right ]
\nonumber\\
     && \cdot \frac{-i }{q^2 + i\varepsilon}\left ( g^{\mu\nu} - \alpha \frac{q^\mu q^\nu}{q^2 + i\varepsilon} \right ),
\end{eqnarray}
The integration over $q^+$ and $q_\perp$ can be performed with $\delta$-functions.
Performing the $q^-$-integral by a contour integral and using the pole  of the heavy quark propagator
at $q^- = (v\cdot p -v^-q^+)/v^+$ we have for the gauge part:
\begin{eqnarray}
\phi_\pm (k^+, k_\perp) \vert_{2c, \alpha}  &=&  -\frac{ \alpha \alpha_s }{2\pi^2} C_F \theta (-q^+)
 \frac{ v\cdot p }{ v^+ \left ( \frac{q^+}{(v^+)^2 }( q^+ - p^+) + q^2_\perp \right )^2 },
\nonumber\\
   k^+ &=& p^+ - q^+, \ \ \ \vec k_\perp =\vec p_\perp - \vec q_\perp.
\end{eqnarray}
If we use it to calculate the convolution in Eq.(7) to determine $H_{P,K}^{(1)}$, we have to calculate
the integral like:
\begin{equation}
  \int dk^+ d^2 k_\perp f(k^+,k_\perp)\phi_\pm (k^+, k_\perp) \vert_{2c, \alpha}
   \propto  \int^0_{-\infty} d q^+ \int d^2 q_\perp \frac{ f(p^+ -q^+, p_\perp- q_\perp)}
      { \left [(v^+)^2 q_\perp^2 + q^+(q^+ -p^+)\right ]^2 }.
\end{equation}
We denote the integral in the right-hand side as $I$.
If the function is nonzero at $k^+ =p^+$ and $\vec k_\perp = \vec p_\perp$  and
is approaching to zero when $k^+ \to \infty$ and $\vec k_\perp \to \infty$, then we find
that the integral is divergent. The convolution $H_{P,K}^{(0)}\otimes \phi_\pi^{(0)}$ clearly satisfies
the above requirement for the function $f$.
Inspecting the integral one finds that divergence is in the region of $q^+ \to 0$ and $\vec q_\perp \to 0$.
We can regularize the divergence by giving the gluon a mass, or by using
dimensional regularization in which the transverse momentum is in the space with the dimension $2-\epsilon_L$.
The singularity in the integral is easily isolated by subtractions:
\begin{eqnarray}
 I&=& \int^0_{-\infty} d q^+ \int d^2 q_\perp
    \left [ \frac{f(p^+ -q^+, p_\perp- q_\perp)} { \left [(v^+)^2 q_\perp^2 + q^+(q^+ -p^+)\right ]^2 }
     - \frac{f(p^+ -q^+, p_\perp- q_\perp)}
     { \left [ (v^+)^2 q_\perp^2 - q^+ p^+)\right ]^2 } \right ]
\nonumber\\
   &&  +\int^0_{-\infty} d q^+ \int d^2 q_\perp \left [ \frac{ f(p^+ -q^+, p_\perp- q_\perp)}
   { \left [ (v^+)^2 q_\perp^2 - q^+ p^+)\right ]^2 } - \frac{ \theta (q^2_{0\perp}-q^2_\perp)
       \theta( q^+ + q^+_0) f(p^+, p_\perp)}
   { \left [ (v^+)^2 q_\perp^2 - q^+ p^+)\right ]^2 } \right ]
\nonumber\\
   && + f(p^+, p_\perp) \int^0_{-\infty} d q^+ \int d^2 q_\perp
         \frac{ \theta (q^2_{0\perp}-q^2_\perp) \theta( q^+ + q^+_0) }
         {\left [ (v^+)^2 q_\perp^2 - q^+ p^+)\right ]^2 },
\end{eqnarray}
where $q^+_0 >0$. In the above each integral is finite except the last one. The final result will
not depend on the slicing parameter $q_0^+$ and $q^2_{0\perp}$. The last integral is divergent. Using
the integral one can identify the divergent part of the wave function as:
\begin{equation}
\phi_\pm (k^+, k_\perp) \vert_{2c,\alpha} = \frac{ \alpha \alpha_s }{4 \pi}  C_F \delta (k^+ -p^+)
   \delta^2 ( \vec k_\perp -\vec p_\perp) \left (\frac{2}{\epsilon_L} \right )
   + {\rm finite\ terms }.
\end{equation}
Following our above analysis we find that the divergence appears in the momentum region of the
exchanged gluon whose momentum takes the patten $q^\mu \sim (\delta^2, {\mathcal O}(1),
\delta, \delta)$ with $\delta \to 0$. In the following we will only give detailed results
of singularities for the wave functions.
Calculating Fig.2a and Fig.2b we also find that the contributions to $\phi_+$ contain the light-cone singularity:
\begin{eqnarray}
\phi_+ (k^+, k_\perp) \vert_{2a,\alpha} &=& - \frac{ \alpha \alpha_s }{4 \pi}  C_F \delta (k^+ -p^+)
   \delta^2 ( \vec k_\perp -\vec p_\perp) \left (\frac{2}{\epsilon_L} \right )
   + {\rm finite\ terms },
\nonumber\\
\phi_+ (k^+, k_\perp) \vert_{2b,\alpha} &=&  \frac{ \alpha \alpha_s }{4 \pi}  C_F \delta (k^+ -p^+)
   \delta^2 ( \vec k_\perp -\vec p_\perp) \left (\frac{2}{\epsilon_L} \right )
   + {\rm finite\ terms }.
\end{eqnarray}
The gauge part of the contribution from Fig.2d to $\phi_{+,-}$ contains an I.R. singularity regularized by the pole
of $\epsilon_I = 4-d$:
\begin{equation}
\phi_{\pm} (k^+, k_\perp) \vert_{2d,\alpha} =\frac{\alpha \alpha_s}{4 \pi} C_F
 \delta (k^+ -p^+)
   \delta^2 ( \vec k_\perp -\vec p_\perp)\left ( \frac{2}{\epsilon_I} \right )
 + {\rm finite\ terms }.
 \end{equation}
This I.R. singularity will be canceled by I.R. singularities from Fig.3.
\par
It is straightforward to find the contributions from Fig.2a and Fig.2b to $\phi_-$.
It is interesting to note that for the contribution from Fig.2b to $\phi_-$ we find that
the gauge part is finite, while the gauge-independent contribution or the contribution in Feynman
gauge is zero in the limit
$u^-\gg u^-$ simply because $\gamma^+ \gamma\cdot u \to 0$, i.e.,
\begin{equation}
\phi_- (k^+, k_\perp) \vert_{2b}  =  \alpha \left [ {\rm \ finite\ terms \ }\right ] .
\end{equation}
The zero contribution from Fig.2b in Feynman gauge leads to that the evolution equation
for $\phi_-$ will be different than that for $\phi_+$. We come back to this later.
We also find that the gauge part of $\phi_-$ from Fig.2a is finite. Hence, the light-cone singularity
of $\phi_+$ comes from Fig.2a, Fig.2b and Fig.2c, while the light-cone singularity of $\phi_-$ comes only
from Fig.2c.

\par

\begin{figure}[hbt]
\begin{center}
\includegraphics[width=9cm]{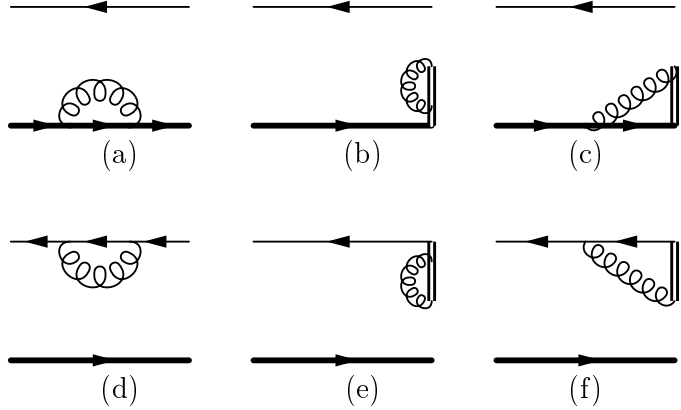}
\end{center}
\caption{The one-loop corrections for the wave functions. }
\label{Feynman-dg1}
\end{figure}
\par
Now we turn the corrections from Fig.3. The contributions from Fig.3a and Fig.3d  do not need to be considered,
because the perturbative coefficients will not receive contributions from them.
It is interesting to look at the contributions to the gauge part of $\phi_{\pm}$ from Fig.3c:
\begin{eqnarray}
\phi_\pm (k^+, k_\perp) \vert_{3c, \alpha} = i\alpha g_s^2 C_F
 \delta (k^+ -p^+)
   \delta^2 ( \vec k_\perp -\vec p_\perp) \int \frac{d^4 q}{(2\pi)^4}
   \frac{v\cdot q}{(q^2+i\varepsilon)^2 (v\cdot (q-p)+i\varepsilon )}.
\end{eqnarray}
This integral can be calculated with the standard method and the result has only
an U.V. divergence.
If we first do the $q^-$-integral as done for Fig.2c, then the remaining integral is proportional
to that in Eq.(15) by setting the function $f$ as $f=1$. In this case, unlike in Eq.(15),
$f$ will not approach to zero
when $q^+ \to -\infty$ or $q_\perp \to \infty$. At first look, the light-cone singularity
with the momentum $q^\mu \sim (\delta^2, {\mathcal O}(1),
\delta, \delta)$ with $\delta \to 0$
appearing in Fig.2c will also appear here. However, in the case $f=1$ one finds another
light-cone singularity with the momentum $q^\mu \sim ( {\mathcal O}(1), \Lambda^2,
\Lambda, \Lambda)$ with $\Lambda \to \infty$. The two light-cone singularities cancel
and it leaves only the U.V. singularity. One can also use arguments from covariance
to show nonexistence of light-cone singularities  in the above integral and in any covariant integral
as shown in \cite{FMW}. We note here that the integral in Eq.(15) with $f$ given as
$H_{P,K}^{(0)} \otimes \phi_\pi^{(0)}$ is not a covariant integral of $q$, because the range of $q^+$
with $-\infty < q^+ \leq 0$  is fixed at the
beginning and $q^-$ is fixed to be zero in $H_{P,K}^{(0)} \otimes \phi_\pi^{(0)}$.
After the U.V. subtraction the gauge parts from Fig.3c are finite. The contributions
from Fig.3b and Fig.3e to the gauge parts have I.R. singularities. The results from Fig.3c, Fig.3b and
Fig.3e are:
\begin{eqnarray}
\phi_\pm (k^+, k_\perp) \vert_{3c, \alpha} &=&  \alpha \left [ \ {\rm finite\ terms }\  \right ],
\nonumber\\
\phi_\pm (k^+, k_\perp) \vert_{3b, \alpha} &=&  \phi_\pm (k^+, k_\perp) \vert_{3e, \alpha}
   = -\frac{\alpha \alpha_s}{8 \pi} C_F \delta (k^+ -p^+)
   \delta^2 ( \vec k_\perp -\vec p_\perp) \left ( \frac{2}{\epsilon_I} \right )
     + {\rm finite\ terms }.
\end{eqnarray}
\par
Similarly, there is no contribution from Fig.3f to $\phi_-$ in Feynman gauge. The gauge part
of $\phi_{+,-}$ has only U.V. divergence. After the U.V. subtraction we have:
\begin{eqnarray}
\phi_+ (k^+, k_\perp) \vert_{3f, \alpha} &=& \alpha \left [ \ {\rm finite\ terms }\  \right ],
\nonumber\\
\phi_- (k^+, k_\perp) \vert_{3f} &=& \alpha \left [ \ {\rm finite\ terms }\  \right ].
\end{eqnarray}
\par
Adding everything together we find that the wave functions contain the light-cone singularity:
\begin{equation}
\phi^{(1)}_\pm (k^+,k_\perp)=\frac{\alpha \alpha_s}{4 \pi} C_F\delta (k^+ -p^+)
   \delta^2 ( \vec k_\perp -\vec p_\perp) \left ( \frac{2}{\epsilon_L} \right )
   + {\rm finite\ terms}.
\end{equation}
The I.R. singularity from Fig.2d is canceled by that from Fig.3b and Fig.3e.
\par
We turn to the wave function of $\pi$. For $\pi$ with the momentum $K^\mu =(0, K^-, 0,0)$
one can define the wave function  with the gauge link along the direction $\tilde u$
with $\tilde u^\mu=(\tilde u^+,\tilde u^-,0,0)$. The wave function is defined
in the limit $\tilde u^+ \gg \tilde u^-$ as
\begin{eqnarray}
\phi_{\pi}(x, k_\perp,\tilde \zeta, \mu) &=& \ \int \frac{ d z^+ }{2\pi}
  \frac {d^2 z_\perp}{(2\pi )^2}  e^{ik^-z^+ - i \vec z_\perp\cdot \vec
k_\perp}
\langle 0 \vert \bar q(0) L_{\tilde u}^\dagger (\infty, 0)
  \gamma^- \gamma_5 L_{\tilde u} (\infty,z) q(z) \vert \pi(K) \rangle\vert_{z^-=0},
\nonumber\\
   k^- &=& xK^-, \ \ \ \ \  \tilde\zeta^2 = \frac{2 \tilde u^+ (K^-)^2}{\tilde u-}\approx
\frac{ 4 (\tilde u\cdot K)^2}{\tilde u^2}.
\end{eqnarray}
We take a quark pair to replace the pion and calculate the wave function.
The quark $q$ of the pair carries the momentum $k_1$ and the antiquark $\bar q$ carries
the momentum $K-k_1$. $k_1$ is given as $k_1^\mu =(0,k_1^-, \vec k_{1\perp})$
with $k_1^-= x_0 K^-$.
At leading order we have:
\begin{equation}
\phi_{\pi}^{(0)}(x, k_\perp) = \delta (k^- -k_1^-) \delta (\vec k_\perp -\vec k_{1\perp}).
\end{equation}
At one-loop level the wave function receives corrections from the same diagrams
in Fig.2 and Fig.3. by replacing the $b$-quark line with the light quark line.
Calculating them in the general covariant gauge, we also find the light-cone singularities. From our
previous results in \cite{FMW} converted with the dimensional regularization we have the gauge parts
containing the light-cone singularity:
\begin{eqnarray}
\phi_\pi (k^-,k_\perp)\vert_{2a,\alpha} &=& -\frac{\alpha \alpha_s}{4 \pi} C_F\delta (k^- -k_1^-)
   \delta^2 ( \vec k_\perp -\vec k_{1\perp}) \left ( \frac{2}{\epsilon_L} \right )
   + {\rm finite\ terms},
\nonumber\\
\phi_\pi (k^-,k_\perp)\vert_{2b,\alpha} &=& +\frac{\alpha \alpha_s}{4 \pi} C_F\delta (k^- -k_1^-)
   \delta^2 ( \vec k_\perp -\vec k_{1\perp}) \left ( \frac{2}{\epsilon_L} \right )
   + {\rm finite\ terms},
   \nonumber\\
\phi_\pi (k^-,k_\perp)\vert_{2c,\alpha} &=& +\frac{\alpha \alpha_s}{4 \pi} C_F\delta (k^- -k_1^-)
   \delta^2 ( \vec k_\perp -\vec k_{1\perp}) \left ( \frac{2}{\epsilon_L} \right )
   + {\rm finite\ terms},
\nonumber\\
\phi^{(1)}_\pi (k^-,k_\perp) &=& \frac{\alpha \alpha_s}{4 \pi} C_F\delta (k^- -k_1^-)
   \delta^2 ( \vec k_\perp -\vec k_{1\perp}) \left ( \frac{2}{\epsilon_L} \right )
   + {\rm finite\ terms}.
\end{eqnarray}
\par
To extract the one-loop perturbative coefficients one needs to use the same quark pairs, used
to calculate the wave functions, to calculate the form factors. At one-loop level, the form factors
will not have such light-cone singularities, as we meet in wave functions.
The reason for this is similar to the case of Fig.3c, where the loop integral is covariant.
With the argument from the covariance one can show that scattering amplitudes, i.e., the form factors,
do not have the light-cone singularity in the general covariant gauge. Details can be found in \cite{FMW}.
This is in agreement with the expectation that any scattering amplitude has only U.V. divergences,
I.R. divergences and collinear divergences.
Actually, scattering amplitudes with off-shell partons will not have any soft divergence like
collinear- and I.R. singularity, because they are all regularized by the off-shellness of partons.
Hence
the one loop perturbative coefficients will have the light-cone singularities:
\begin{eqnarray}
H_P^{(1)} (x, \vec p_\perp, y, \vec k_\perp ) &=& - \frac{\alpha \alpha_s}{4 \pi} H_P^{(0)} (x, \vec p_\perp, y, \vec k_\perp )
\left [ \left ( \frac{2}{\epsilon_L} \right ) +
\left ( \frac{2}{\epsilon_L} \right ) \right ] +{\rm finite\ terms},
\nonumber\\
H_K^{(1)} (x, \vec p_\perp, y, \vec k_\perp ) &=& - \frac{\alpha \alpha_s}{4 \pi} H_K^{(0)} (x, \vec p_\perp, y, \vec k_\perp )
\left [ \left ( \frac{2}{\epsilon_L} \right ) +
\left ( \frac{2}{\epsilon_L} \right ) \right ] +{\rm finite\ terms}.
\end{eqnarray}
It should be noted that the singular contributions depend on gauges. Therefore, the $k_T$-factorization
is gauge dependent. Also because it is singular, the $k_T$-factorization can not hold beyond tree-level in
the general covariant gauge.
\par
Now we turn to the $\zeta$-dependence of $\phi_{\pm}$. This dependence can be used to do resummation of large log's.
The dependence of $\phi_+$ can be found in \cite{MWWF}, where it is determined with on-shell quark pair
and in Feynman gauge. From our calculation one can verify that the dependence
is gauge independent. The result for $\phi_+$ is\cite{MWWF}:
\begin{eqnarray}
\zeta \frac{\partial}{\partial \zeta} \phi_+(k^+,b,\mu, \zeta ) &=&
   \left [-\frac{4\alpha_s}{3\pi} \ln\frac{\zeta^2 b^2 e^{2\gamma-1}}{4}
       -\frac{2\alpha_s}{3\pi} \ln\frac{\mu^2e}{\zeta^2} \right ]
       \phi_+(k^+,b,\mu, \zeta)
\nonumber\\
  &=& \left [ K(\mu,b) + G(\mu,\zeta)
      -\frac{2\alpha_s}{3\pi} \ln\frac{\mu^2e}{\zeta^2} \right ] \phi_+(k^+,b,\mu, \zeta),
\end{eqnarray}
it should be noted that the kernel is not exactly the famous factor $K+G$,
the last factor comes because we used HQET for the heavy quark.
In the above the wave function is defined in the impact space as:
\begin{equation}
\phi_\pm(k^+, b,\mu, \zeta) = \int d^2 k_\perp e^{i \vec k_\perp \cdot \vec b}
       \phi_\pm (k^+,k_\perp, \mu, \zeta) .
\end{equation}
\par
The $\zeta$-dependence is introduced by those diagrams in Fig.2 and Fig.3, where the exchanged
gluon is emitted or absorbed by the gauge link. Only from Fig.2b and Fig.3f,
$\phi_+$ receives contributions which are proportional to $\ln^2\zeta^2$. This results in that
Eq.(28) contains a single log of $\zeta^2$. This dependence is useful for resummation
of Sudkov double log's.
As we already noticed before that Fig.2b and Fig.3f
do not contribute to $\phi_-$ in Feynman gauge. Except these two diagrams, other diagrams
contribute to the same $\zeta$-dependence to $\phi_- $ as to $\phi_+$. Using our previous results
in \cite{MWWF} we find the evolution equation for $\phi_-$:
\begin{eqnarray}
\zeta \frac{\partial}{\partial \zeta} \phi_-(k^+,b, \mu, \zeta ) =
   \left [-\frac{2\alpha_s}{3\pi} \ln\frac{\mu^2 b^2 e^{2\gamma}}{4} \right ]
       \phi_-(k^+,b,\mu, \zeta).
\end{eqnarray}
From the above, the $\zeta$-dependence of $\phi_-$ is different than that of $\phi_+$. Therefore
we in general can not have $\phi_+ =\phi_-$.
The $\zeta$-dependence of $\phi_-$
is a constant of $\zeta$, it can only be used to re-sum single log's.
By using the solutions of the above evolution equations one can resum possible log terms
in perturbative coefficients into an exponential factor $S$. Since the evolution equations
are different, the exponential factor $S$ related to $\phi_+$ or the form factor $F_P$ is
different than that related to $\phi_-$ or the form factor $F_K$. This implies that
the resummation in the $k_T$-factorization with only one wave function in \cite{KT1,KTB1,KTB2}
can not be consistent.
\par
To our conclusion: The $k_T$-factorization has been widely used for exclusive $B$-decays
and it is quite successful in phenomenology. However, its successfulness does not
imply theoretical consistence of the factorization.
In the $k_T$-factorization of exclusive $B$-decays the perturbative
coefficients are extracted from scattering amplitudes of off-shell partons. Because
the scattering amplitudes are gauge-dependent, one can not expect in general that
the extracted perturbative coefficients are gauge invariant. We have taken
the semi-leptonic decay $B\to \pi \ell \nu_\ell$ as an example to show that the perturbative coefficients
are indeed gauge-dependent. They contain light-cone singularities
in the general covariant gauge. The singularities disappear in Feynman gauge.
The singular contributions come only from wave functions. This leads to the conclusion
that the $k_T$-factorization of exclusive $B$-decays is gauge dependent and violated
in the general covariant gauge because of the singularities.
A gauge-dependent factorization is generally unacceptable since it deliveries
physical predictions which are gauge-dependent.
The $k_T$-factorization
employs only one wave function to describe the nonperturbative property of a $B$-meson.
In fact there are two wave functions. The two wave functions are different and they satisfy
different evolution equations which can be used to re-sum large log's. Because the difference
of the two wave functions and their evolution equations, the resummation of large log's
at leading log approximation in
the $k_T$-factorization with only one wave function
is inconsistent.

\par\vskip20pt
\par\noindent
{\bf\large Acknowledgments}
\par
This work is supported by National Nature Science Foundation of P.R.
China(No.10721063, 10575126, 10747140).
\par


\end{document}